\documentclass[iop,apj]{emulateapj}

\usepackage{natbib}

\shorttitle{GRB 090902B: afterglow observations and implications}

\slugcomment{Accepted to ApJ}

\begin{document}

\title{GRB 090902B: afterglow observations and implications}

\author{S.~B.~Pandey\altaffilmark{1,2}, 
C.~A.~Swenson\altaffilmark{3}, 
D.~A.~Perley\altaffilmark{4}, 
C.~Guidorzi\altaffilmark{5}, 
K.~Wiersema\altaffilmark{6}, 
D.~Malesani\altaffilmark{7}, 
C.~Akerlof\altaffilmark{1}, 
M.~C.~B.~Ashley\altaffilmark{8}, 
D.~Bersier\altaffilmark{9}, 
Z.~Cano\altaffilmark{9}, 
A.~Gomboc\altaffilmark{10}, 
I.~Ilyin\altaffilmark{11}, 
P.~Jakobsson\altaffilmark{12}, 
I.~K.~W.~Kleiser\altaffilmark{4}, 
S.~Kobayashi\altaffilmark{9}, 
C.~Kouveliotou\altaffilmark{13}, 
A.~J.~Levan\altaffilmark{14}, 
T.~A.~McKay\altaffilmark{1}, 
A.~Melandri\altaffilmark{9}, 
C.~J.~Mottram\altaffilmark{9}, 
C.~G.~Mundell\altaffilmark{9}, 
P.~T.~O'Brien\altaffilmark{6}, 
A.~Phillips\altaffilmark{8}, 
J.~M.~Rex\altaffilmark{4}, 
M.~H.~Siegel\altaffilmark{9}, 
R.~J.~Smith\altaffilmark{9}, 
I.~A.~Steele\altaffilmark{9}, 
G.~Stratta\altaffilmark{15}, 
N.~R.~Tanvir\altaffilmark{6}, 
D.~Weights\altaffilmark{16},  
S.~A.~Yost\altaffilmark{17}, 
F.~Yuan\altaffilmark{1}, 
W.~Zheng\altaffilmark{1}}

\altaffiltext{1} {Randall Laboratory of Physics, Univ. of Michigan, 450 Church Street, Ann Arbor, MI, 48109-1040, USA}
\altaffiltext{2} {Aryabhatta Research Institute of Observational Sciences, Manora Peak, Nainital, India, 263129}
\altaffiltext{3} {Pennsylvania State Univ., 525 Davey Lab, University Park, PA 16802, USA}
\altaffiltext{4} {Dept. of Astronomy, Univ. of California, Berkeley, CA 94720-3411, USA}
\altaffiltext{5} {Dipartimento di Fisica, Universit{\`a} di Ferrara, via Saragat 1, I-44100, Ferrara, Italy}
\altaffiltext{6} {Dept. of Physics and Astronomy, Univ. of Leicester, University Road, Leicester, LE1 7RH, UK}
\altaffiltext{7} {Dark Cosmology Centre, Niels Bohr Institute, Univ. of Copenhagen, Juliane Maries vej 30, DK-2100 Copenhagen \O, Denmark}
\altaffiltext{8} {School of Physics, Univ. of New South Wales, Sydney NSW 2052, Australia}
\altaffiltext{9} {Astrophysics Research Institute, Liverpool JMU, Twelve Quays House, Egerton Wharf, Birkenhead, CH41 1LD, UK}
\altaffiltext{10} {Faculty of Mathematics and Physics, Univ. of Ljubljana, Jadranska 19, SI-1000 Ljubljana, Slovenia}
\altaffiltext{11} {Astrophysikalisches Institut Potsdam, An der Sternwarte 16, D-14482 Potsdam, Germany}
\altaffiltext{12} {Centre for Astrophysics and Cosmology, Science Institute, Univ. of Iceland, Dunhagi 5, IS-107 Reykjav\'ik, Iceland}
\altaffiltext{13} {NASA, Marshall Space Flight Center, NSSTC, 320 Sparkman Drive, Huntsville, Alabama, 35805, USA}
\altaffiltext{14} {Dept. of Physics, Univ. of Warwick, Coventry, CV4 7AL, UK}
\altaffiltext{15} {ASI SDC, via Galileo Galilei, 00044 Frascati, Italy}
\altaffiltext{16} {Dept. of physical sciences, Univ. of Hertfordshire, College Lane, Hatfield, AL10 9AB, UK}
\altaffiltext{17} {Dept. of Physics, College of St. Benedict, Collegeville, MN 56321, USA}

\shortauthors{Pandey S. B. et al. 2010}

\begin{abstract}

The optical-infrared afterglow of the LAT-detected long duration burst, GRB 090902B, has been observed by several 
instruments. The earliest detection by ROTSE-IIIa occurred 80 minutes after detection by the GBM instrument onboard 
the {\it Fermi} Gamma-Ray Space Telescope, revealing a bright afterglow and a decay slope suggestive of a reverse shock 
origin. Subsequent optical-IR observations followed the light curve for 6.5 days. The temporal and spectral behavior 
at optical-infrared frequencies is consistent with synchrotron fireball model predictions; the cooling break lies between 
optical and XRT frequencies $\sim$ 1.9 days after the burst. The inferred electron energy index is $p = 1.8 \pm 0.2$, 
which would however imply an X-ray decay slope flatter than observed. The XRT and LAT data have similar spectral indices 
and the observed steeper value of the LAT temporal index is marginally consistent with the predicted temporal decay in 
the radiative regime of the forward shock model. Absence of a jet break during the first 6 days implies a 
collimation-corrected $\gamma$-ray energy $E_{\gamma} > 2.2\times10^{52}\rm$ ergs, one of the highest ever seen in 
a long-duration GRBs. More events combining GeV photon emission with multi-wavelength observations will be required 
to constrain the nature of the central engine powering these energetic explosions and to explore 
the correlations between energetic quanta and afterglow emission.

\end{abstract}


\section{Introduction}

The recently launched {\it Fermi} Gamma-Ray Space Telescope with the on-board Gamma-ray Burst 
Monitor (GBM) and Large Area Telescope (LAT) instruments \citep{atwood09, meegan09} 
in conjunction with the {\it Swift} narrow field instruments \citep{gehrels04} have opened a 
new window to understand the physical mechanisms that generate GeV photons in very energetic GRBs 
and the relation to lower energy components of the afterglow \citep{band09}. Since the {\it Fermi} 
launch more than a year ago, only 14 GRBs have been detected by the LAT while more than $\sim$ 350 bursts 
have been seen by the GBM during the same period. Optical afterglows have been detected for 7 of the 14 
LAT events starting from $\sim$ 300 s to a few hours after the burst. The origin of these high-energy photons 
and their possible correlation to afterglow emission is still debated \citep[and references therein]{zou09}. 
GRB 080916C \citep{abdo09b}, GRB 090510 \citep{abdo09c, max10} and GRB 090902B \citep{abdo09a} are 
among the brightest LAT bursts indicating some signatures 
consistent with the synchrotron forward shock models \citep{kumar09a, kumar09b, ghirlanda10}.
However, in the case of GRB 090902B, the deviation of the burst spectrum from the Band function 
and the observed large amplitude variability at very short time-scales \citep{abdo09a} does not 
support the afterglow origin of the LAT data. High energy photons from GRBs have previously been observed by 
the EGRET detector and has shown evidence for deviations from synchrotron models \citep{hurley94, gonzalez03}. 

The bright GRB 090902B (trigger 273582310) was detected by the GBM on 2009
Sep 2nd at 11:05:08.31 UT with an initial error box radius of 2-3 degrees centered at R.A. 
= 17${^h}$ 38${^m}$ 26${^s}$, Dec. = +26$\degr$ 30$\arcmin$ and a burst duration 
of 21.9 s in the energy band  50 - 300 keV \citep{bissaldi09}.
The burst was one of the brightest at LAT energies with a power-law spectral distribution
{at both} low and high energies \citep{depalma09}. The detailed analysis of the LAT and GBM data 
has been presented in \citet{abdo09a}. The observed features in the prompt burst spectrum have also 
shown evidence for an underlying photospheric thermal emission \citep{ryde10}. ToO observations with 
{\it Swift} started $\sim$ 12.5 hours after the GBM trigger. The X-ray afterglow was detected within 
the LAT error-circle by the XRT \citep{kennea09}, the UVOT \citep{swenson09} and later 
by several other ground-based multiwavelength facilities. The burst redshift, $z$ = 1.822, was determined by the 
Gemini-North telescope \citep{cucchiara09}. The afterglow was also
seen at radio frequencies by the WSRT \citep{vander09} and by the VLA \citep{chandra09}.

The details of the optical-infrared (IR) observations along with the temporal and spectral properties 
of the afterglow are described in the next section. In \S 3, we discuss the observed 
properties of the afterglow and comparisons to various models. These results are summarized in
\S 4. Throughout the paper, we use the usual power-law representation of flux density,
$f{_\nu}(t) \propto \nu^{-\beta} t^{-\alpha}$, for the regions without spectral breaks
where $\alpha$ and $\beta$ are the power-law temporal decay and spectral indices, respectively. 
For the cosmological calculations we have used the cosmological parameters 
$H_0=71~\rm km~s^{-1}Mpc^{-1}$, $\Omega_{M}=0.27$, $\Omega_{\Lambda}=0.73$. 
Errors are quoted at the 1-sigma level unless otherwise stated.

\section{Observations and Data Reduction}

The receipt of a ground-corrected GBM trigger with a 1 degree nominal location error initiated an observing 
sequence for ROTSE-IIIa, located at the Siding Spring Observatory in Australia. The telescope began 
taking three sets of thirty 20-s images, tiled around the GBM estimated location. Only the third set, 
starting 80 minutes after the burst, with R.A. = 17${^h}$ 38${^m}$ 13${^s}$ and Dec. = +27$\degr$ 30$\arcmin$ 59$\arcsec$,
covered the XRT burst location later identified as R.A. = 17${^h}$ 39${^m}$ 45.26${^s}$ and Dec. = +27$\degr$ 19$\arcmin$ 28.1$\arcsec$ 
\citep{kennea09}. A substantial fraction of the delay was imposed by the overlap of a 
previous observation request for an unrelated field. Since the burst occurred during the early afternoon 
in Namibia and Turkey, ROTSE-IIIc and d could not respond until 6 hours later at which point similar 
sequences of images were obtained. The raw images were processed using the standard ROTSE software pipeline
and photometry was performed on co-added images using the method described in \citet{quimby06}. 
The ROTSE-IIIa observations were taken under bright night sky 
conditions with a lunar phase two days short of full. By the time the relevant exposures began, the 
transient was $13\,^{\circ}$ above the horizon, decreasing to less than $11\,^{\circ}$ at completion. 

Because of the difficult observing environment, we carefully examined various possibilities that could 
lead to a false identification. All such efforts were confined to a 329$\times$304 pixel sub-image 
spanning 18$\arcmin \times$17$\arcmin$, centered at the OT. Within this field, 34 false positives were 
identified by SExtractor with signatures similar to the GRB OT. After correcting for 
the small fraction of the image too bright for such detections, the probability of such an event within a 
5$\times$5 pixel region close to the GRB coordinates established by later observations is less than 1\%.
We next examined the sequence of 30 images carefully to determine if the apparent OT signal was an 
artifact of one or two frames with spurious problems. Five USNO stars with $13.2 \leq m_R \leq 14.3$ 
were chosen 
to establish the image point spread function and the sky extinction, frame-by-frame. Not surprisingly, 
the sky extinction and the ambient sky brightness increased by 30\% and 20\%, respectively, over the 890 
second duration of these observations. For each image, the OT amplitude was determined and the entire set 
was fit to a power-law in time, constrained to the observation obtained by the Nickel telescope \citep{perley09}  
17 hours post-burst. The best fit corresponds to $m_R$ = 16.4$\pm$0.5 at $t = 5320$ s after the burst and
located within 1$\arcsec$ of later, deeper detections. Based on the 
statistical analysis of the total ensemble of 30 measurements, we estimate a spurious detection probability 
of less than 1\%. Combined with the spatial localization constraint, the probability of a false 
identification is less than 1$\times10^{-4}$.

The UVOT magnitudes in $u$ band were calculated using the UVOT photometric system \citep{poole08} and the XRT data were 
reduced using the standard tools \citep{evans07}. The $V, R$ and $I$ magnitudes of the afterglow from the data 
taken by 1.0m Nickel, 2.0m Liverpool and 2.5m NOT telescopes were computed using nearby stars in the GRB field 
calibrated on October 10 under good photometric sky conditions by the 2.5m NOT telescope. 
The $r'$ magnitudes were obtained using GROND \citep[for details][]{mcbreen10, olivares09}. 
The $J$ and $K$ magnitudes of the afterglow were computed with respect to nearby 
2MASS stars from the data taken by 3.8m UKIRT and 4.2m WHT. 
The photometry of the optical afterglow from these observations are summarized in Table 1 and the $V, R, I$ 
magnitudes of the 5 nearby stars in the GRB field are given in Table 2.

\begin{table*}
{\bf Table 1.}~Observations of the afterglow of GRB~090902B at optical-IR frequencies by the consortium of telescopes. \\
\begin{center}
\begin{tabular}{ccccc} 
\tableline\tableline \hline
Time since GRB &  Filter & Telescope    & Exposure time  & Magnitude         \\
(s)&       &         &   (s)      &     \\
\tableline \hline
4803         & $R$     & 0.45m ROTSE-IIIa & 890      & 16.4$\pm$0.5 \\
23245        & $R$     & 0.45m ROTSE-IIId & 846      & $>$18.7        \\
25438        & $R$     & 0.45m ROTSE-IIIc & 892      & $>$18.6        \\
\hline
62291        & $R$     & 1.0m Nickel     & 5$\times$600+300 & 20.60$\pm$0.10 \\
74749        & $R$     & 2.0m Liverpool & 2$\times$900      & 21.04$\pm$0.11 \\
123605       & $R$     & 2.0m Liverpool & 2$\times$900       & 21.40$\pm$0.10  \\
299554       & $R$     & 2.0m Liverpool & 2$\times$900      & 22.60$\pm$0.25   \\
328027       & $R$     & 2.0m Liverpool & 2$\times$900      & $>$22.1          \\
\hline
153674       & $i'$     & 2.0m Liverpool & 2$\times$900    &  21.33$\pm$0.15    \\
330039       & $i'$     & 2.0m Liverpool & 2$\times$900    & $>$21.5 \\
\hline
132935       & $r'_{AB}$    & 2.2m GROND     & 738       & 21.54$\pm$0.05   \\
218021       & $r'_{AB}$    & 2.2m GROND     & 738       & 22.01$\pm$0.07   \\
563787       & $r'_{AB}$    & 2.2m GROND     & 738       & 23.07$\pm$0.17   \\
\hline
135552       & $I$    & 2.5m NOT     & 3$\times$300       & 20.72$\pm$0.11   \\
134499       & $R$    & 2.5m NOT     & 3$\times$300       & 21.40$\pm$0.11   \\
133462       & $V$    & 2.5m NOT     & 3$\times$300       & 21.67$\pm$0.11   \\
\hline
164490       & $J$     & 3.8m {\it UKIRT-WFCAM} & 1080   & 20.20$\pm$0.20 \\
164780       & $K$     & 3.8m {\it UKIRT-WFCAM} & 1080   & 18.90$\pm$0.25 \\
\hline
122083       & $J$     & 4.2m {\it WHT-LIRIS} & 24$\times$75   & 19.99$\pm$0.15 \\
125366       & $K$     & 4.2m {\it WHT-LIRIS} & 48$\times$54   & 18.92$\pm$0.20 \\
\hline
45097        & $u$     & {\it Swift}-UVOT & 1075   & 20.34$\pm$0.18    \\
50891        & $u$     & {\it Swift}-UVOT & 1616   & 20.63$\pm$0.17    \\
56674        & $u$     & {\it Swift}-UVOT & 1997   & 20.52$\pm$0.14    \\
62458        & $u$     & {\it Swift}-UVOT & 2589   & 21.11$\pm$0.20    \\
68244        & $u$     & {\it Swift}-UVOT & 2515   & 20.90$\pm$0.16    \\
73990        & $u$     & {\it Swift}-UVOT & 2552   & 21.01$\pm$0.18 \\
79775        & $u$     & {\it Swift}-UVOT & 1311   & 20.65$\pm$0.21 \\
104201       & $u$     & {\it Swift}-UVOT & 3853   & 21.92$\pm$0.35 \\
312211       & $u$     & {\it Swift}-UVOT & 10549  & 22.27$\pm$0.20 \\
664679       & $u$     & {\it Swift}-UVOT & 9969   & $>$ 22.8 \\
\tableline
\end{tabular}
\end{center}
\end{table*}



\begin{table}
{\bf Table 2.}~The $(\alpha_{2000}, \delta_{2000})$ of the 5 stars near the afterglow position and
their standard magnitudes in $V$, $R$ and $I$ photometric passbands. 
\begin{center}
\tiny
\begin{tabular}{ccc cc ccl} \hline \hline
ID & $\alpha_{2000}$ & $\delta_{2000}$ & $V$ & $R$ & $I$ \\
    & $({\rm h:m:s})$ & $(\circ:\prime:\prime\prime)$ & (mag) & (mag) & (mag)  \\ \hline \hline
 A&17 39 48.8&+27 19 57.0& 18.98$\pm$0.02& 18.38$\pm$0.01&  17.92$\pm$0.01 \\
 B&17 39 44.8&+27 19 26.0& 18.67$\pm$0.01& 18.35$\pm$0.01&  17.93$\pm$0.02 \\
 C&17 39 42.7&+27 19 13.9& 17.76$\pm$0.01& 17.34$\pm$0.01&  16.99$\pm$0.01 \\
 D&17 39 47.6&+27 18 46.7& 17.79$\pm$0.01& 17.39$\pm$0.01&  17.02$\pm$0.01 \\
 E&17 39 49.2&+27 18 53.6& 18.84$\pm$0.01& 18.57$\pm$0.02&  18.14$\pm$0.02 \\
\hline
\end{tabular}
\end{center}
\end{table}


\begin{figure}
  \includegraphics[scale=0.42]{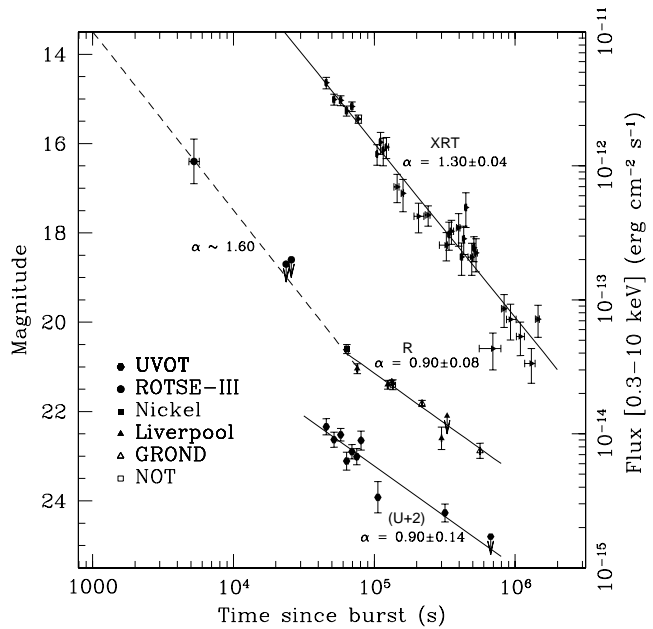}
  \caption{X-ray and optical afterglow light curves of the GRB 090902B. The solid lines 
mark the best fit power-laws to the XRT, $u$ and $R$ band light curves. An offset of 2 
magnitudes has been applied to the $u-$band data for clarity. The power-law segment 
between the first ROTSE data point and the Nickel data is shown by a dashed line.}
\end{figure}

\subsection{Afterglow Light Curves and Spectral Energy Distribution}

The optical data for the afterglow taken in $R, r'$ and $u$ filters along with 
the publicly available XRT data are plotted in Figure 1. Single power-law fits to 
the $R$ and $u$ band data $>$ 12.5 hours post-burst have temporal decay indices of 
$0.90 \pm 0.08$ and $0.90 \pm 0.14$ respectively. To obtain continuity, the ROTSE optical 
data require a steeper decline at earlier times, with a temporal decay index 
$\sim 1.6$ or greater. The XRT light curve is characterized by a single power-law with a decay index 
of $1.30 \pm 0.04$, starting from 12.5 hours to 17 days after the burst.

The {\it Swift}-XRT time averaged spectrum ($\sim$ 12.5 to 413 hours post burst) 
has been analyzed using XSPEC with an absorbed power-law model and $z=1.822$, 
inferring a spectral index $\beta_X = 0.9\pm0.1$ and a rest-frame column density 
$N_H^{z}=(1.8\pm0.3)\times10^{22}$ cm$^{-2}$ in addition to the Galactic column density
$N_H^{G}=(3.8\pm0.3)\times10^{20}$ cm$^{-2}$ in the direction of the burst.
The spectral analysis of the XRT data between $\sim$ 12.5 and 20 hours post burst determines 
a power-law index of $\beta_X = 1.0\pm0.1$ (and $N_H^{z}=(2.4\pm0.4)\times10^{22}$ cm$^{-2}$).
For the period of $\sim$ 20 to 413 hours, the comparable value is $\beta_X = 0.75\pm0.25$ 
(and $N_H^{z}=(0.7\pm0.4)\times10^{22}$ cm$^{-2}$), indicating no significant spectral evolution
during these observations. In subsequent analysis, a value of $0.9\pm0.1$  
will be assumed for $\beta_X$. The afterglow spectral energy distribution (SED) is constructed at $\sim$ 1.9 days 
post-burst using optical-IR data at $K, J, I, r', R, V$ and $u$ bands along with the XRT data 
as shown in Figure 2. The optical-IR spectral index, after correcting only for the Galactic extinction 
$E(B-V)$ = 0.04 mag \citep{schlegel98}, has $\beta_O = 0.68\pm0.11$, flatter than that measured at 
XRT frequencies. 

\section{Results and Discussions}

\begin{figure}[t]
\vspace{-1.3cm}
  \includegraphics[scale=0.45]{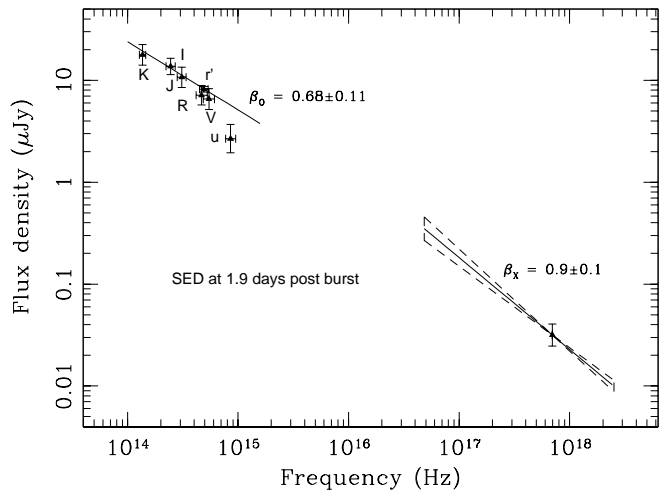}
  \caption{Multi-wavelength spectral energy distribution of the GRB 090902B afterglow at optical-IR and XRT frequencies 
derived at 1.9 days post-burst. The epoch has been chosen to allow the best possible spectral coverage.
The observations in the $u$ band might be effected by Ly-$\alpha$ and hence were excluded to determine the 
spectral index $\beta_O$.
}
\end{figure}

\subsection{The Optical Brightness and LAT-Detected Bursts}

The power-law decays seen in other early optical afterglows of GRBs \citep{panaitescu08, oates09b} 
suggest that the single observed data point at $\sim$ 1.4 hours is unlikely to 
be a flaring feature at such late times. For an observation $\sim$ 1.4 hours after the burst, the ROTSE 
detection at $m_R \sim$ 16.4 mag is remarkably bright. This is best qualified by the statistical study of a 
large ensemble of bursts afterglows by \citet{akerlof07}. They found that the temporal evolution of 
the brightness distribution is well described by a power-law exponent, $\alpha \sim$ 0.7. 
With that behavior, a magnitude of 16.4 at 1.4 hours would have evolved from 
$m_R \sim$ 15.2 at t = 1000s and thus lie among the top 5 \% of all bursts. 
Even one-magnitude errors in the ROTSE measurement would not substantially 
modify this conclusion. At later times, the optical afterglow must drop with 
a much steeper slope of $>$ 1.6. If this behavior was manifested earlier, the 
brightness of GRB 090902B was even more pronounced. 

The apparent temporal decay index, $\alpha >$ 1.6, between 1.4 to 12.5 hours is steeper 
than the value of $\alpha = 0.90 \pm 0.08$ for epochs $>$ 12.5 hours by more than 7-$\sigma$ and is 
consistent with the dominance of reverse shock origin \citep{sari99a, kobayashi00, zhang03} as seen recently 
for the energetic `naked eye' GRB 080319B, on similar time scales \citep[e.g.] [] {bloom09, pandey09b}. 

Comparison of the observed apparent optical brightness of GRB 090902B at 
$\sim$ 1.4 hours to a much larger sample of pre-{\it Swift} and {\it Swift} optical afterglows
\citep[See Figure 1] [] {kann07} also indicates that the GRB 090902B was one of the brightest at 
such early epochs. A handful of other examples of long-duration GRBs, detected by LAT and with measured redshift 
values (GRB 080916C \citep{greiner09}; GRB 090323 \citep{updike09}; GRB 090328 \citep{oates09a}; 
GRB 090926 \citep{haislip09} and GRB 091003 \citep{gronwall09} have been observed at optical 
frequencies starting $\sim$ 16 to 26 hours post-burst. Comparison of the observed apparent optical 
brightness of the LAT-detected bursts 16-26 hours post-burst to the sample published in \citet{kann07} indicates 
that the optical brightness of LAT-detected GRBs are typical except GRB 090926 \citep{haislip09} which was 
one of the brightest even at $\sim$ 20.0 hours post-burst. Thus, the late time behavior of LAT-detected GRBs at 
optical frequencies is not unusual.

\subsection{Afterglow Models and GRB 090902B}

The derived values of temporal and spectral indices from multi-wavelength data can be
compared with the closure relations \citep{price02} to discriminate between interstellar 
medium (ISM) and wind ambient profiles \citep{sari98, chevalier00} and 
to infer the location of the cooling break, $\nu_c$. For the observed values of $\alpha$ ($>$ 12.5 hours) 
and $\beta_O$ at optical frequencies, the closure relation $\alpha = 3\beta/2$ is satisfied within errors 
in case of the ISM model \citep{price02} for the observed frequencies $\nu < \nu_{c}$. Also, the value 
of the temporal decay index at XRT frequencies is steeper than for the optical, which clearly 
rules out the wind model and requires the ISM model with $\nu_{c}$ between optical and XRT frequencies. 
The location of $\nu_{c}$ below XRT frequencies implies that the electron energy index is $p = 1.8\pm0.2$, 
deduced solely using the value of $\beta_X$. The value of $p$ and 
the determined temporal slopes at optical frequencies are also consistent with the closure relation 
$\alpha = 3(p-1)/4$ within errors, valid for the spectral regime $\nu_{m} < \nu < \nu_{c}$ in case of the 
ISM model. However, the observed temporal decay index at XRT frequencies is inconsistent with the ISM model 
closure relation $\alpha = (3p-2)/4$ (for $\nu > \nu_{c}$) by 2.8$\sigma$ and requires a steeper value of $p$ 
than estimated. The predicted value of $p$ for $\nu > \nu_{c}$ is 2.4$\pm$0.05 using the XRT temporal decay 
index $\alpha$ = 1.30$\pm$0.04. The afterglow properties favor an evolution of $\nu_{c}$ between the XRT 
and optical frequencies during the observations with an expected $\beta_O$ of $0.4\pm0.1$ for $p = 1.8\pm0.2$.
The relatively shallower value of $\beta_O$ than observed can be attributed to a moderate amount 
of extinction $A_V$ = 0.20$\pm$0.06 mag for SMC-like dust and assuming $\leq$ 20\% of the $u$ flux 
is affected by Ly-$\alpha$ at the SED epoch using the method described in \citet{perley08}. 
The present optical-IR and XRT data have determined the value of $\nu_c$ that is contrary to the assumption 
of \citep{kumar09b} that $\nu_{c}$ lies above the XRT frequencies and thus implies a steeper value of $p$. 

In the light of above discussions, the published radio data at 4.8 GHz \citep{vander09} and 
8.46 GHz \citep{chandra09} of GRB 090902B near the SED epoch was used to constrain 
location of the self absorption frequency $\nu_a$. The expected value of the 
spectral index between 4.8 and 8.46 GHz will be $\sim$ -0.4, closer to the expected $\nu^{-1/3}$ spectral regime for 
$\nu_a < \nu_m$ in the case of slow-cooling forward shock model \citep{sari98}. The effect of 
scintillation has not been taken into account which might modify the flux values for the observed frequencies 
at early epochs. Using the observed flux values at the radio frequencies and assuming $\nu_m$ to be 
$< 5.0\times10^{14}$ Hz at $\sim 10^{4}$ s after the burst, the estimated value of the peak synchrotron flux 
at the SED epoch is $\sim$ 0.5 milliJy. The values of the peak synchrotron flux and $\nu_m$ at the epoch of SED 
are used to constrain the value of $\nu_a$ using equation 4.9 of \citet{sari01}. The calculated value of
$\nu_a < 10^{8}$ Hz is below the observed radio frequencies and in agreement with the slow-cooling model for 
$\nu_a < \nu_m$ at the epoch of the SED.

The analysis also indicates no signature of a possible jet-break before or during the period
of our afterglow observations. For the measured fluence between 10 keV and 10 GeV
\citep{abdo09a}, the inferred value of the isotropic equivalent energy is 
$E_{\gamma}^{iso\rm}=3.6\times10^{54}\rm$ ergs 
assuming a gamma-ray efficiency $\eta_\gamma =$ 0.2 and the circumburst density $n =$ 1 cm$^{-3}$
\citep{frail01}. Based on the observed properties of the burst, if we 
limit the jet-break time to be greater than 6 days after the burst, the value of the jet opening angle is $\theta_j >
0.11$ rad which gives the collimation corrected energy $E_{\gamma} > 2.2\times10^{52}\rm$ ergs, one of 
the highest ever inferred \citep{cenko10}. With known values of $p$, the measured XRT flux 
at 1 day after the burst and using the description given in \citet{freedman01}, the isotropic 
fireball energy carried by electrons is $\epsilon_eE = 3.1\times10^{54}\rm$ ergs, where $\epsilon_e$ is 
the fraction of shock energy carried by relativistic electrons, comparable to $E_{\gamma}^{iso\rm}$ \citep[see also]
[]{starling09, tanvir09}. The constraint on the energetics of the burst is comparable to the 
energy budget in the case of magnetars \citep{usov92, starling09, cenko10} and 
could also be accommodated within the ``collapsar'' origin of GRBs \citep{mac99}. The
energy estimates of more LAT-detected GRBs in the future will help towards a 
better understanding the nature of the central engine powering these energetic events. 

\subsection{Onset of the GeV Afterglow}

The detection of many delayed photons at energies $>$ 1 GeV, the observed high value of the isotropic $\gamma$-ray 
energy and the very early peak time seen in the LAT light curve constrain the value of the bulk Lorentz 
factor $\Gamma$ to be $\sim 1000$ \citep{abdo09a}. Such high values of $\Gamma$ have also been estimated 
in the case of other LAT bursts, GRB 080916C \citep{abdo09b, greiner09} and GRB 090510 
\citep{abdo09c, ghirlanda10}. Along with the afterglow properties discussed in the previous 
sections, the very high value of $\Gamma$ and the very early peak in the LAT light curve provide a good 
opportunity to test the LAT temporal decay and spectral properties \citep{abdo09a} for an early onset 
of the afterglow in terms of synchrotron shock models \citep{sari98, sari99b}. 
 
Under the synchrotron fireball model, a power-law distribution can be assumed in both time and spectral domains. 
Based on the discussions in the previous section and assuming $\nu_c \sim 2\times10^{16}$ Hz at the epoch of SED, 
the extrapolated values 
of $\nu_m$ and $\nu_c$ at 100.0 s are $< 5.0\times10^{17}$ Hz and $\sim 8\times10^{17}$ Hz respectively (assuming a 
temporal scaling of $\nu_c \propto t^{-1/2}$ and $\nu_m \propto t^{-3/2}$), both below 10 keV. At the SED epoch, 
the observed flux density of 0.03$\pm$0.01~$\mu$Jy (at 2.88 keV) will give rise to an extrapolated flux density
of 450$\pm$150~$\mu$Jy at 100.0 s assuming the temporal decay of $\sim$ 1.3. The 
1 GeV flux density calculated at 100.0 s \citep{abdo09a} is $\sim$ 0.004 $\mu$Jy. These 
flux densities at 1 GeV and 2.88 keV imply a spectral index of $\sim$ 0.9 at 100.0 s, in
agreement with the XRT spectral index at the epoch of SED and the LAT spectral index within 2-$\sigma$ 
\citep{abdo09a, ghisellini10}. This indicates that for GRB 090902B, both XRT and LAT frequencies share the same 
spectral regime with $\nu_c$ below XRT frequencies under the synchrotron model, although the temporal index of 
$\sim$ 1.5, observed at LAT frequencies is steeper than the XRT temporal decay index 1.3$\pm$0.04. However, 
our results show that around 100.0 s, $\nu_m < \nu_c$ and the observed temporal decay index at LAT frequencies is 
marginally consistent with the expected temporal decay index of (2 - 6p)/7 in the radiative case of the synchrotron 
model \citep{sari98}. Recently, based on the bolometric afterglow luminosity estimates for radiative 
fireballs, the expected temporal decay index $t^{10/7}$ for the LAT frequencies \citep{ghisellini10} is also 
close to the observed LAT temporal index of  $\sim$ 1.5.

In the case of another LAT-detected GRB 080916C, the value of spectral index at LAT frequencies is $1.1\pm0.1$ 
\citep{ghisellini10} and the value of GeV flux density at 100.0 s is $\sim$ 0.006 $\mu$Jy \citep{abdo09b}. 
Using the XRT data analysis published in \citep{greiner09}, the extrapolated value of 2.88 keV flux density at 
100.0 s is 250$\pm$50~$\mu$Jy. For GRB 080916C, the spectral index between 2.88 keV and 1 GeV at 100.0 s comes out 
to be $\sim$ 0.9, close to the spectral index seen at the LAT frequencies \citep{ghisellini10}. This also
indicates that $\nu_c$ is between XRT and LAT frequencies at 100.0 s within the assumptions of the afterglow model 
proposed by \citep{greiner09}. In the case of GRB 090510, the multi-wavelength SED at 100.0 s 
after the burst supports the afterglow origin of LAT data in terms of synchrotron forward shock model with a possible 
energy injection at optical and XRT frequencies \citep{max10}. The observed values of XRT flux 
for GRB 080916C, GRB 090510 and GRB 090902B are typical for other well observed Swift 
GRBs\footnote[1]{http://www.swift.ac.uk/xrt\_curves} \citep{zheng09} at similar time scales.

Based on above discussion, the evidence in favor of synchrotron forward 
shock model for the observed GeV emission for GRB 090902B, GRB 090510 and 
GRB 080916C is consistent with the predictions made by \citep{zou09}.  
However, the hard photon index at LAT frequencies and the evidence for reverse 
shock emission in early optical data cannot rule out the possibilities of 
synchrotron self-Compton emission at LAT frequencies \citep{wang01} 
and the Klein-Nishina suppression of high energy electrons at early times 
\citep{wang10}. Such processes would require theoretical modeling 
which is beyond the scope of this paper.

\section{Summary}

We present the observations of the afterglow of GRB 090902B at optical-IR frequencies carried out 
from 80 min to 6.5 days after the burst. The comparison of the optical afterglow of GRB 090902B to 
other bright pre-{\it Swift} and {\it Swift} burst indicates that the optical afterglow was bright at 
early epochs but decreased at times to a level typical of other bursts. The apparently steeper temporal 
decay of the early optical data can be explained in terms of reverse shock emission based on the estimated 
value of the early peak time provided by the reverse shock emission model \citep{zhang03}. 
These inferred parameters are also in agreement with the very early onset 
of the afterglow seen at LAT frequencies \citep{abdo09a}. The temporal and spectral decay nature at 
optical-IR frequencies at later epochs favor synchrotron forward shock model although temporal decay at XRT 
frequencies requires a steeper electron energy index than the deduced value of $p$ = 1.8$\pm$0.2 solely from 
the XRT spectral index and $\nu_c$ between optical and XRT frequencies. The radio afterglow data constrains 
self-absorption frequency $\nu_a < \nu_m$ which is lower than the observed radio frequencies at the SED epoch. 
The LAT and XRT data of GRB 090902B share similar spectral slopes at early epochs and indicate towards their 
common origin under the synchrotron forward shock model for the radiative fireballs. Also, the present analysis 
can not rule out the possible non-synchrotron origin for the emission at LAT frequencies. The estimated value of 
$E_{\gamma}$ along with the required amount of host extinction is consistent with a massive star origin for the burst. 
However, the delayed GeV emission and the delayed onset are commonly observed properties of both long and short-duration 
GRBs detected by the LAT \citep{abdo09a, abdo09b, abdo09c} in spite of differences in many observed properties 
including the proposed differences for their progenitors. More LAT-detected GRBs, especially seen by the BAT as well,
and their early follow-up observations using ground-based robotic telescopes will shed light on the 
temporal properties of the afterglows and their possible correlation with the observed GeV emission. Finally, the 
most tantalizing aspect of these observations is the realization that a great deal more could be learned if 
accurate LAT localizations could be made available more promptly.

\vspace{0.1cm}

\acknowledgments

This research has made use of the data obtained through the High Energy Astrophysics Science 
Archive Research Center Online service, provided by the NASA/Goddard Space Flight Center. 
The ROTSE project is supported by the NASA grant NNX08AV63G and the NSF grant PHY-0801007. 
The authors associated with the WHT and the UKIRT acknowledge the Isaac Newton Group of 
Telescopes at La Palma and the Science and Technology Facilities Council of UK respectively. 
DARK is funded by DNRF. We also thank Gabor Furesz for his observations with the NOT.

\end{document}